\documentclass[
aps,
prl,
groupedaddress,
superscriptaddress,
floatfix,
twocolumn
]{revtex4-1}

\usepackage{graphicx}
\usepackage{xcolor}
\usepackage{amsmath,amssymb}
\usepackage{hyperref} 
\usepackage{bm} 
\usepackage{mathtools} 
\usepackage{float}
\usepackage{braket}

\newif\iffigures
\figurestrue

\newcommand{\p}{\partial}

\begin{document}

\title{Chiral solitons in spinor polariton rings}

\author{D. A. Zezyulin}
\affiliation{ITMO University, St. Petersburg 197101, Russia}

\author{D. R. Gulevich} 
\affiliation{ITMO University, St. Petersburg 197101, Russia}

\author{D. V. Skryabin} 
\affiliation{ITMO University, St. Petersburg 197101, Russia}
\affiliation{Department of Physics, University of Bath, Bath BA2 7AY, United Kingdom}

\author{I. A. Shelykh}
\affiliation{ITMO University, St. Petersburg 197101, Russia}
\affiliation{Science Institute, University of Iceland, Dunhagi 3, IS-107, Reykjavik, Iceland}

\date{\today}

\begin{abstract} 
We consider theoretically one-dimensional polariton ring accounting for both longitudinal- transverse (TE-TM) and Zeeman splitting of spinor polariton states and spin dependent polariton-polariton interactions. We present the novel class of solutions in the form of the localized defects rotating with constant angular velocity and analyze their properties for realistic values of the parameters of the system. We show that the effects of the geometric phase arising from the interplay between external magnetic field and TE-TM splitting introduce chirality in the system and make solitons propagating in clockwise and anticlockwise directions non equivalent. This can be interpreted as solitonic analog of Aharonov-Bohm effect. 
\end{abstract}

\maketitle


\textit{Introduction.} Topology of the potential is critical for the dynamics of a quantum particle, since it defines  connectivity of the  available trajectories. Therefore changes in topology of a system are often related to qualitative alternations of its physical behaviour \cite{book}.  Quantum non-single connected structures, such as mesoscopic rings, reveal a rich variety of quantum mechanical effects \cite{review1,review2,review3}. One prominent example is the famous Aharonov-Bohm effect \cite{Siday,Aharonov-Bohm}, where the phase of a charged particle is influenced by the magnetic field,  which is effectively zero at the particle's location. This results in magnetic-flux-dependent oscillations of the ring-confined particle energy and of the conductivity of the system in the ballistic regime.

For neutral particles with spin, an analog of the Aharonov-Bohm phase is represented by the geometric Berry phase. The latter appears if an effective magnetic field responsible for the energy splitting of the two components of a spinor changes smoothly it the direction along the ring. In the adiabatic approximation when spin follows the direction of the magnetic field the phase acquired by a particle during one cycle of the propagation along the ring is equal to half of the solid angle covered by the vector of an effective magnetic field. The geometric phase was experimentally detected in photonic interferometers \cite{Tomita,Kwiat} and predicted to play a substantial role in excitonic \cite{Moulopoulos} and polaritonic \cite{Shelykh2005} ring resonators. The latter system will be in focus of our attention in the present work.

Cavity polaritons are hybrid light-matter quasiparticles emerging in the regime of the strong coupling between excitonic resonance and photonic mode of planar semiconductor microcavity \cite{Hui}. Compared to purely photonic or purely excitonic systems polaritonic systems have sevaral important advantages. From their photonic component polaritons inherit extremely small effective mass (about $10^{-5}$ of the mass of free electrons) and large coherence length (in the mm scale) \cite{Ballarini2017}. On the other hand, presence of the excitonic component results in polariton-polariton interactions, which can be controlled  by means of external electromagnetic fields \cite{Device}. 

An important property of cavity polaritons is their spin (or pseudo-spin) \cite{ShelykhPSSb},
inherited from the spins of QW excitons and cavity photons. Similar to photons, polarions have two possible spin projections on the structure growth axis corresponding to two opposite circular polarizations. States with opposite spins can be mixed by effective magnetic fields of various origin.  Magnetic field applied along the structure growth axis and acting on excitonic component splits  energies of the spin positive and spin negative polariton states, while TE-TM splitting of the photonic modes hybridizes these states via the linear coupling\cite{ShelykhPSSb}. 
Because of the effective spin-orbit interaction it provides to polaritonic systems~\cite{Kavokin-PRL-2005, Leyder-NP-2007, Sala-PRX-2015}, TE-TM splitting has recently been shown to play an important role in various types of phenomena in artificial lattices~\cite{Jacqmin-PRL-2014, Kusudo-2013, Milicevic-2015, Kibble-Zurek, Gulevich-kagome, Bleu-Hall, Gulevich-SciRep-2017, Bleu-valley, Gulevich-Yudin, Lieb}.
    
\begin{figure}
	\begin{center}
		\includegraphics[width=1.0\columnwidth]{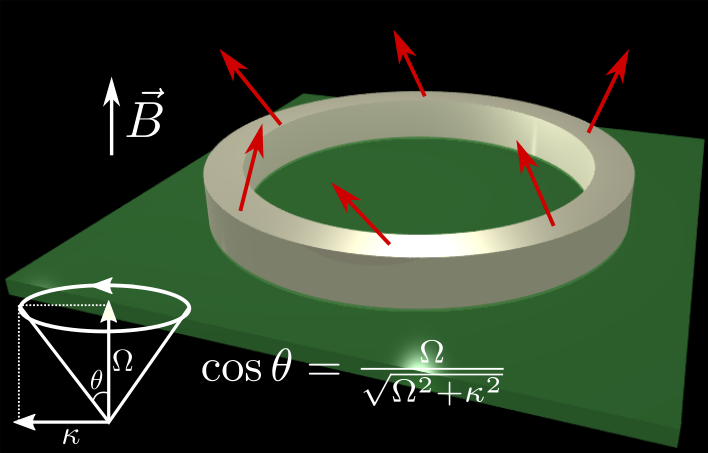}
	\end{center}
	\caption{Schematic illustration of the considered geometry. Polariton ring is placed into external magnetic field ${\bf B}$ perpendicular to its interface. The total effective magnetic field acting on polariton's spin is a combination  of the real magnetic field and the field provided by TE-TM splitting. The direction of the total effective magnetic field changes along the ring as it is shown by the red arrows. If one moves along the ring it covers a cone  characterized by the angle $\theta$.} 
	\label{figure1}
\end{figure}

\textcolor{black}{The polariton  interactions render the system   nonlinear and enable propagation of self-sustained nonlinear entities (i.e., solitons) whose properties depend  significantly on the underlying topology.  The importance of topological solitons has been known in several sub-areas of nonlinear field theory including nonlinear optics and cold atom physics,
see, e.g. \cite{DS1, DS2}. Their robustness is topologically protected that makes them both attractive for potential applications
and readily observable in experiments even in the presence of unavoidable dissipation. Effects of balancing between
pump and loss have also been studied in the context of topological localised structures, see, e.g. \cite{DS3}. Recently \textit{chiral}
effects in nonlinear spinor field models have attracted attention in the context of information processing in both quasi-conservative, gain and
pump free, systems \cite{DS4, DS5, DS6} as well as in the dissipative models with pump \cite{DS7}.}
In our recent paper \cite{Gulevich2016}, we analyzed how the combination of the effects of the geometric phase and spin-dependent polariton-polariton interactions affects stationary nonlinear states in the polariton rings. On the other hand, it is well known that 1D polariton systems support wide variety of propagating topological defects \cite{Sich2016} including solitons and half solitons \cite{Flayac2011}, analogs of magnetic monopoles \cite{Hivet2016}, propagating domain walls \cite{Liew2008} and others. The goal of the present letter is to analyze \textit{rotating} nonlinear solutions in 1D spinor polariton rings \textcolor{black}{which can be readily realised in practical devices \cite{DS8}}.

\begin{figure}
 {\includegraphics[width=1.0\columnwidth]{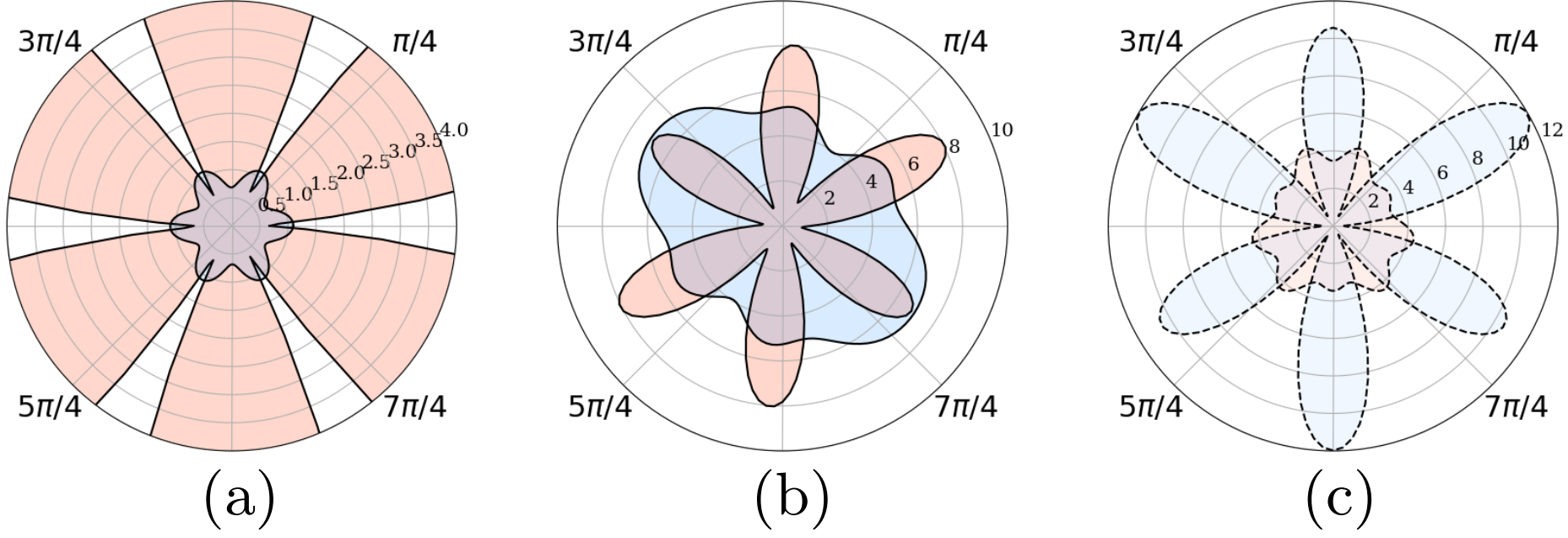}}
\caption{\label{fig:j3-classes} 
Different types of rotating solitons  at
$\alpha=-0.05$, $\rho=10$, $\kappa=0.8$, $\Omega=0.5$, $\omega=0.4$.
The pink (blue) color denotes the density of the $\psi_+$ ($\psi_-$)  component. Saturated colors with solid lines (a,b) and dull colors with dashed lines (c)  mark the linearly stable and unstable configurations, respectively.}
\end{figure}

\textit{The model.} Interacting spinor polaritons trapped in a quasi one-dimensional ring resonator (see Fig.~\ref{figure1}) can be described by the following system of dimensionless Gross-Pitaevskii equations \cite{Gulevich2016}:
\begin{equation}
\begin{array}{l}
i\dot\psi_\pm= - \partial_\varphi^2\psi_\pm + \left( |\psi_\pm|^2 + \alpha |\psi_\mp|^2 \right)\psi_\pm \quad\quad \\[3mm]
\hspace{3cm}
\pm\Omega\psi_\pm  + \kappa e^{\mp2 i \varphi} \psi_\mp,
 \end{array}
\label{GP-ring}
\end{equation}
Here, $\psi_{\pm}$ are the components of the  exciton-polariton spinor wavefunction $\pmb\psi\equiv\{\psi_+,\psi_-\}$ in the basis of circular polarizations satisfying $\psi_{\pm}(t,\varphi)=\psi_{\pm}(t,\varphi+2\pi)$, parameter $\alpha<0$ characterizes attractive interaction of the cross-polarized polaritons, $\Omega$ is half of the Zeeman energy splitting proportional to the amplitude of the applied magnetic field and $\kappa$ is half of the momentum independent TE-TM energy splitting. Parameters $\Omega$ and $\kappa$ are dimensionless and scale in units of $\hbar^2/(2 m^* R^2)$, where $R$ is the ring radius  and $m^*$ is the exciton-polariton effective mass. 
\textcolor{black}{Unit energy $\hbar^2/(2m^* R^2)$ can be varied in a broad range. 
Depending on the detuning between the photon and exciton frequencies, it can take values from 4 $\rm \mu eV$ to 40 $\rm \mu eV$ for ring radius 5 $\rm \mu m$~\cite{Gulevich2016}. 
TE-TM splitting can be made both as high as $\sim 1 meV$ in a waveguide of width 1~$\rm\mu m$ \cite{Kuther,Dashbach}
and negligibly small by choosing large ring widths. Also, TE-TM splitting can be controlled by changing detuning~\cite{Duff-2015} 
and properties of distributed Bragg reflector~\cite{Panzarini}. With the unit energy 40 $\rm \mu eV$ and angular velocity $\omega\sim 1$
it will take about 100 picoseconds for a soliton to circle around the ring, which is of the order of polariton lifetime.
Normalized densities $\psi_\pm$ scale out the polariton interaction constant (which can be of order $\sim 10^{-5}$meVmm$^{-2}$ \cite{DZ1}, and coefficient $\alpha$ is the ratio of interactions between polaritons with parallel and antiparallel spins, which is a small negative constant  \cite{DZ1} (it can also be controlled by the detuning  between exciton 
and photon modes \cite{DZ2}). System \eqref{GP-ring} describes quasi-conservative nonlinear dynamics of exciton-polaritons which
has been observed in several experiments \cite{DS9,DS10}. It is expected that the unavoidable (small) losses will limit the lifetime of  solitons, but will not inhibit their chiral properties discussed in what follows.}

To find  rotating solutions we switch from the laboratory frame to the frame of reference rotating with frequency $\omega$, which is achieved by replacing the polar angle $\varphi$ with a new variable $x = \varphi-\omega t$. Seeking 
the wave functions in the form $\psi_{\pm}(t,\varphi) = 
u_{\pm}(t,x) e^{-i\mu t \mp i \varphi}$,
where $u_\pm(t,x)$ satisfy periodic boundary conditions: $u_\pm(t, x) = u_\pm(t, x+2\pi)$, we get the following system:
\begin{equation}
i\dot u_\pm = ( \hat D_{\pm} +|u_\pm|^2 + \alpha |u_\mp|^2 ) u_\pm    + \kappa u_\mp,
\label{GP-phi}
 \end{equation}
where $\hat D_{\pm}= -\mu + i\omega\p_x -(\p_x \mp i)^2 \pm \Omega$.
Nonlinear spinor systems similar to that in Eqs.~(\ref{GP-phi}) have been considered earlier in the context of birefringent optical fibers  \cite{Malomed91,book2} and, more recently, for modelling of matter waves in spin-orbit coupled Bose-Einstein condensates \textcolor{black}{ \cite{SO-BEC,SO-BEC-new}}. 
Chemical potential $\mu$ and frequency $\omega$ uniquely select 
the polariton density integral $\rho\equiv \frac{1}{2\pi}\int_0^{2\pi}\left(|\psi_+|^2+|\psi_-|^2\right)d\varphi$ for a given  solution.

\begin{figure}
 {\includegraphics[width=1.0\columnwidth]{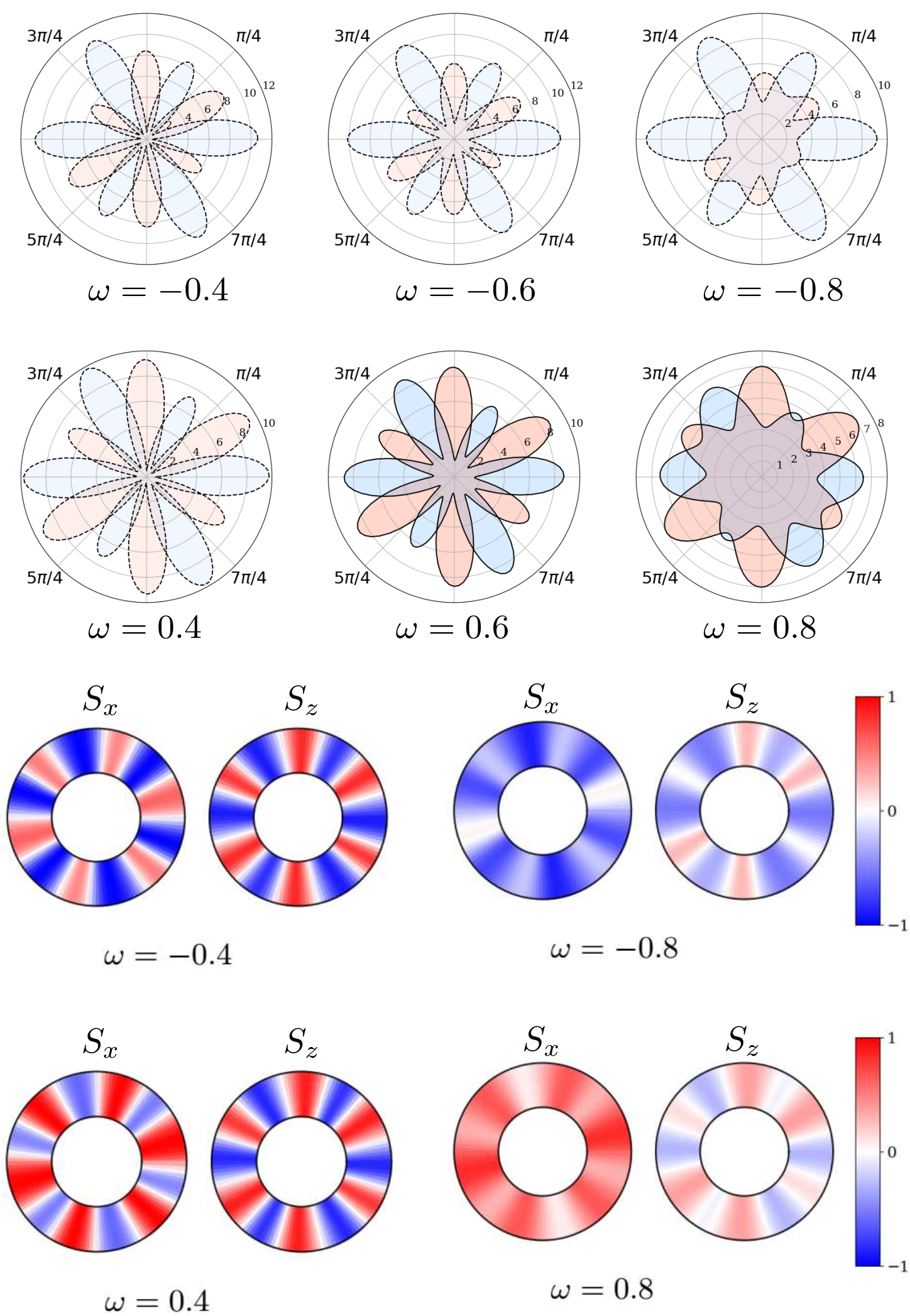}}
\caption{\label{fig:j3} 
Six upper panels: several representatives of a family of solutions of type (iii) with  $\alpha=-0.05$, $\rho=10$, $\kappa=0.8$, and $\Omega=0.5$ at different angular velocities $\omega$.
Unstable solutions are marked by dull colors and dashed lines. Linearly stable configurations are marked by saturated colors and solid lines. Lower panels show pseudocolor visualizations of Stokes parameters $S_x$ and $S_z$ for  $\omega=\pm 0.4$ and $\omega=\pm 0.8$.
}
\end{figure}

We compute  rotating solutions by the numerical continuation  from the analytically tractable limit $\kappa=\Omega=\omega=\alpha=0$, where equations \eqref{GP-phi} decouple into a pair of nonlinear Schr\"odinger equations (NLSEs) whose 
solutions can be found  in terms of Jacobi elliptic functions \cite{Carr-I}. Proceeding in this way, we have discovered that   system \eqref{GP-phi}  supports  a  rich variety of nonlinear rotating patterns. We mention here three classes of those.
First, \textcolor{black}{when the initial condensate state is chosen to be periodic in one component and exactly zero in the second component, the numerical iterative procedure converges to a solution} 
such that the amplitude of one component is much larger than that of the other one, i.e., either $|\psi_+| \gg |\psi_-|$ or  $|\psi_-| \gg |\psi_+|$, see Fig.~\ref{fig:j3-classes}(a). Since the smaller component can be neglected, the properties of this family can be recovered from the scalar NLSE equation. 
\textcolor{black}{Starting from an initial state with a periodic profile in one component and constant and nonzero density in another one, we obtain class of solutions where}
one of the components has strongly modulated amplitude with pronounced density humps and dips, while the other one is modulated relatively weakly, see Fig.~\ref{fig:j3-classes}(b,c).
In the third family,  which is in the focus of our attention here, \textcolor{black}{initial states with nontrivial periodic densities in both $\psi_+$ and $\psi_-$ lead to solutions where }
both components feature strong density modulation, see \textcolor{black}{Fig.~\ref{fig:j3}}.  The simplest  solution like this features two density dips and two density humps. In general, they can have  arbitrarily large, but always even  number of petals. Every density dip in this class of solutions corresponds to the $\pi$ phase shift and therefore an even number of them is required  to satisfy the periodic boundary conditions.  In what follows, we illustrate the main results of our study using the solutions with six petals. However, we have checked that our observations and conclusions  also remain valid for smaller and larger number of petals.

\begin{figure}
  {\includegraphics[width=0.8\columnwidth]{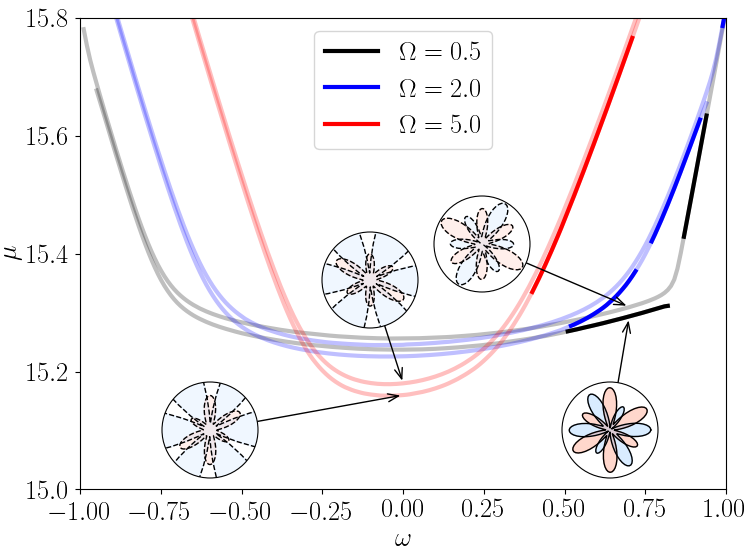}}
\caption{\label{fig:mu} 
Dependencies of the chemical potential $\mu$ on the rotation velocity $\omega$ at different values of the magnetic field $\Omega$. Several configurations of solitons are shown in insets.
For each set of parameters, two families of rotating  solutions  
arise which differ by the phase difference between the two circularly polarized components. Fragments with saturated and dull colors correspond to stable and unstable solutions, respectively. Here   $\alpha=-0.05$, $\rho=10$, and $\kappa=0.8$.}
\end{figure}

\textit{Chiral solitons.}  Now  we proceed to  the \textit{main results} of our work.
With the   focus  on  rotating patterns, it is  of  obvious interest to investigate if and how their properties   are affected by the angular velocity $\omega$. To answer this question, in the six upper panels of Fig.~\ref{fig:j3} we show  representative solutions of the third class with six dip-hump pairs. Solutions in different panels differ by their  angular velocities $\omega$. 
One   interesting feature immediately visible in these panels  is that the increase of the magnitude  of velocity  can be favorable for solution's stability: for instance, the unstable solution with $\omega=0.4$ becomes stable as the velocity increases   to $\omega=0.6$ and, further, to $\omega=0.8$. Even more interestingly, from Fig.~\ref{fig:j3}  we observe that solitons propagating with  opposite velocities ($\omega$ and $-\omega$) have essentially different shapes and, generally speaking, different stability properties. For instance,  the unstable solution rotating in the counter-clockwise direction with  velocity $\omega=-0.6$ becomes stable as the rotation's direction is switched to clockwise  ($\omega=0.6$).  This means that the found solitons are inherently \textit{chiral}, in the sense that  solitons propagating with angular velocities of equal amplitudes but opposite directions are not equivalent. In order to highlight additionally the  differences in the structure of solitons propagating with opposite velocities, in lower panels of  Fig.~\ref{fig:j3} we  compare pseudocolor plots of Stokes parameters defining the distribution of the linear and circular polarization degree along the ring for two  pairs of  counter-propagating patterns:
\begin{equation*}
S_x = \frac{2\textrm{Re}(\psi_+^*\psi_-)}{|\psi_+|^2 + |\psi_-|^2}, \quad S_z = \frac{|\psi_+|^2 - |\psi_-|^2}{|\psi_+|^2 + |\psi_-|^2}.
\end{equation*}

\begin{figure}
\includegraphics[width=0.8\columnwidth]{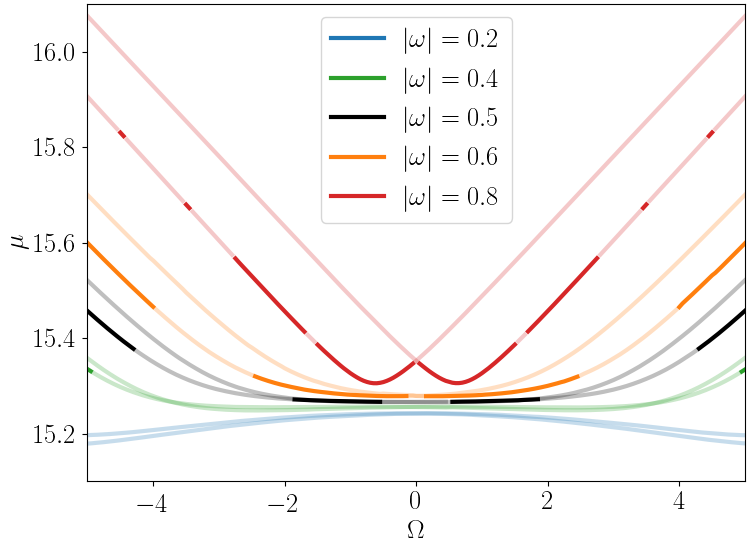}
\caption{\label{fig:Omega} 
Dependencies of the chemical potential on the magnetic field $\Omega$ for solitons in Fig.~\ref{fig:mu}. 
Fragments with saturated and dull colors correspond to stable and unstable solutions, respectively. 
}
\end{figure}

To demonstrate that the revealed chirality does not depend on the particular choice of the parameters, in the lower panel of Fig.~\ref{fig:mu} we plot several  dependencies of the chemical potential $\mu$ on the rotation velocity $\omega$ at different nonzero values of the magnetic field $\Omega$. For each value of $\Omega$, the upper line and lower lines are two different types solutions which differ by phases of the two circular polarized components (cf. symmetric and antisymmetric constant amplitude solutions in \cite{Gulevich2016}). All the obtained dependencies are not symmetric with respect to the vertical axis $\omega=0$ and consist of  unevenly distributed subfamilies of stable and unstable solutions, which is another clear evidence of the solitons' chirality.     Additionally, in Fig.~\ref{fig:Omega} we show the dependence of the chemical potential $\mu$ on the magnetic field $\Omega$ for solitons in Fig.~\ref{fig:mu}. For every considered velocity, the curves with $\omega$ and $-\omega$ intersect exactly at $\Omega=0$.
As another remarkable feature,  we note that  at rotation velocities $\omega=\pm 0.5$ the solitons exhibit topological spin-Meissner effect~\cite{Gulevich2016}
in the region $|\Omega| \lesssim 2$ when the energies are almost independent of the magnetic field.

To understand  the 
the role of external magnetic field, we notice that if the Zeeman splitting is absent [i.e., $\Omega=0$ in (\ref{GP-phi})], then a soliton propagating with velocity $\omega$ always has a twin propagating with the opposite velocity $-\omega$. Both these counter-propagating  solitons   have identical properties, i.e., one can be rendered to another by inverting the spatial direction from $x$ to $-x$ and swapping the polarizations -- these transformations obviously do not alter the physical properties of the solution. In the meantime, nonzero 
$\Omega$ 
breaks this symmetry and   implies that solitons propagating with two opposite velocities $\omega$ and $-\omega$ 
have different properties. However, if    reversing   the velocity is accompanied by  the change  of the magnetic field from $\Omega$ to $-\Omega$, then the system becomes invariant under an evident symmetry: indeed, if $u_+(x)$ and $u_-(x)$ are time-independent solutions of equations  (\ref{GP-phi}), then a pair of new functions $U_+(x) = u_-(-x)$ and $U_-(x) = u_+(-x)$  solves the same system with $\omega\to-\omega$, $\Omega\to-\Omega$.

The chiral nature of solitons is linked with breaking of the time inversion symmetry which appears naturally in the systems with present gauge fields. It can be shown that in the system we consider synthetic U(1) gauge field appears due to the combination of TE- TM splitting and external magnetic field acting on polariton spin. Indeed, let us consider the projection of the original system of the equations onto the lowest energy spin state in adiabatic approximation \cite{Dalibard11}. Diagonalizing the Hamiltonian associated with system (\ref{GP-ring}) in the basis of dressed states, one observes that in the limit of small densities $|\psi_\pm|\ll 1$ the adiabatic dynamics  is governed by the following effective density-dependent Hamiltonian \cite{Malomed17}:
$\hat{H} = (\hat{p}-A)^2 +g|\psi_1|^2$,
where $\hat{p}=-i\partial_\varphi$ is the momentum operator, $|\psi_1|^2$ is the local   squared density of $\psi_+$-component in one of the dressed states,  $g$ is the  effective nonlinearity coefficient \cite{SM}: 
\begin{equation}
g = \frac{2\Omega^2 +\kappa^2(\alpha+1)}{\Lambda(\Lambda + \Omega)}  - \frac{(1-\alpha)\kappa^2\Omega}{\Lambda^4},
\end{equation}
where $\Lambda = \sqrt{\kappa^2 + \Omega^2}$. The geometric density-dependent gauge field reads
\begin{equation}
\label{eq:A}
A = ({\Omega}\,{\Lambda}^{-1} -1)  + (1-\alpha)\,{\Omega}\,\Lambda^{-2} (1-{\Omega}\,{\Lambda^{-1}})|\psi_1|^2,
\end{equation}

The appearance of the gauge field can be qualitatively explained in the following way. Suppose that polariton is moving adiabatically along the ring. If spin-dependent polariton-polariton interactions are neglected, the direction of the total effective magnetic field acting on polaritons spin changes along the ring according to the   formula ${\bf B}={\bf e}_x\kappa \cos2\varphi+{\bf e}_y\kappa \sin2\varphi+{\bf e}_z\Omega$ (see Fig.~\ref{figure1}). 
In adiabatic approximation spin follows the direction of the magnetic field, and therefore when polariton completes one round of the propagation along the ring its spin covers non-zero solid angle which leads to the appearance of the geometric phase equivalent to 
\begin{equation}
2\pi(\cos\theta-1)=2\pi(\Omega/\Lambda-1)=\int_{0}^{2\pi}Ad\varphi. 
\end{equation}
From this expression one immediately deduces the first term in Eq.~(\ref{eq:A}) corresponding to the linear regime.

\textit{To conclude}, in this work we have introduced a novel class of solitons which have the form of localized defects rotating with constant angular velocity in a spinor polariton ring. 
The properties of the solitons, such as the spatial shape and the dynamical stability, can be effectively managed by the angular velocity. 
The solitons feature    chiral nature which makes the solutions propagating in clockwise and counterclockwise direction not equivalent. The chirality is  explained using the concept of effective gauge field stemming from the combined effect of the TE-TM splitting and the external magnetic field.

\begin{acknowledgments}
\textit{Acknowledgments}. 
We thank Dr. I.V. Iorsh for valuable discussions and preparation of the Figure~1. This work was supported by megagrant 14.Y26.31.0015 and Goszadanie no. 3.8884.2017/8.9 and 3.2614.2017/4.6 of the Ministry of Education and Science of Russian Federation. I.A.S acknowledges support from the Icelandic Research Fund, Grant No. 163082-051. D.V.S. acknowledges support from the Royal Society.
\end{acknowledgments}

\pagebreak
\widetext
\begin{center}
	\textbf{\large Supplemental Material for \textit{Chiral solitons in spinor polariton rings}}
\end{center}
\setcounter{equation}{0}
\setcounter{figure}{0}
\setcounter{table}{0}
\setcounter{page}{1}
\makeatletter
\renewcommand{\theequation}{S\arabic{equation}}
\renewcommand{\thefigure}{S\arabic{figure}}
\renewcommand{\bibnumfmt}[1]{[S#1]}
\renewcommand{\citenumfont}[1]{S#1}

\section*{Adiabatic approximation}

Without loss of generality, we assume $\kappa \geq 0$ in system (1) in the main text.  The case with negative values of $\kappa$   is  equivalent  to inverting the sign of one of the polarizations (e.g. $\psi_- \to -\psi_-$).

In order to consider the adiabatic dynamics induced by nonlinear system (1) in the main text, we notice that
it is  associated with the Hamiltonian  $\hat{H}_0 = \hat{p}^2 + \hat{H}_{lin} + \hat{H}_{int}$, where $\hat{p}=-i\partial_\varphi$ is the momentum operator, and $\hat{H}_{lin}$ and $\hat{H}_{int}$ describe the linear part of the system and the interactions, respectively:
\begin{eqnarray}
H_{lin} = \left( \begin{array}{cc}
\Omega & \kappa e^{-2i\varphi}\\
\kappa e^{2i\varphi} & -\Omega
\end{array} \right), 
\quad
H_{int} = \left( \begin{array}{cc}
|\psi_+|^2 + \alpha |\psi_-|^2& 0\\
0 & |\psi_-|^2 + \alpha |\psi_+|^2
\end{array} \right).
\end{eqnarray}
Eigenvalues of $H_{lin}$ are $\lambda_1^{(0)} = \Lambda$ and $\lambda_2^{(0)} =-\Lambda$, where 
\begin{equation}
\Lambda = \sqrt{\Omega^2 + \kappa^2}.
\end{equation}
The corresponding  orthonormal (for any $\varphi$) eigenvectors read
\begin{equation}
|\chi_1^{(0)}\rangle = \frac{1}{\sqrt{2\Lambda}}\left(\begin{array}{c}
\sqrt{\Lambda+\Omega}\\[2mm]
e^{2i\varphi}\sqrt{\Lambda-\Omega}
\end{array}\right), \quad 
|\chi_2^{(0)}\rangle = \frac{1}{\sqrt{2\Lambda}}\left(\begin{array}{c}
-e^{-2i\varphi}\sqrt{\Lambda-\Omega}\\[2mm]
\sqrt{\Lambda+\Omega}
\end{array}\right).
\end{equation}
In the limit of small densities, i.e., $|\psi_{\pm}| \ll 1$,  $H_{int}$ can be treated as a perturbation to $H_{lin}$. Then the perturbed eigenvalues have the form
\begin{eqnarray}
\lambda_{1,2} = \pm  \Lambda + \langle \chi_{1,2}^{(0)}|H_{int}|\chi_{1,2}^{(0)}\rangle,
\end{eqnarray} 
and the perturbed dressed states are
\begin{eqnarray}
|\chi_{1,2}\rangle =  |\chi_{1,2}^{(0)}\rangle \pm  (2\Lambda)^{-1}  \langle \chi_{2,1}^{(0)}|H_{int}| \chi_{1,2}^{(0)}\rangle \  |\chi_{2,1}^{(0)}\rangle,
\end{eqnarray} 

In order to compute the perturbations,  we notice that the condition of local minimization of the free energy implies that for small densities, i.e., for  $|\psi_\pm|^2 \ll 1$, the densities satisfy the  relation \cite{SRK06S}: $S_z =(\pm\Omega n/2)/\sqrt{\Omega^2 + \kappa^2}$ where $n=|\psi_+|^2 + |\psi_-|^2$ is the total concentration, and $S_z=1/2(|\psi_+|^2 - |\psi_-|^2)$ is the density imbalance. To be specific, we consider the upper sign.  Therefore the densities in the components are related as
$|\psi_+|^2\left(\Lambda - {\Omega}\right) = |\psi_-|^2\left(\Lambda + {\Omega}\right)$.
Then the perturbation expansions can be rewritten in the form
\begin{eqnarray}
\lambda_{1} &=& \phantom{+} \Lambda + \frac{2\Omega^2 +\kappa^2(\alpha+1)}{\Lambda(\Lambda + \Omega)}|\psi_{1}|^2,\\
\lambda_{2} &=&  -\Lambda + \frac{2\alpha\Omega^2 + \kappa^2(\alpha+1)}{\Lambda(\Lambda + \Omega)}|\psi_{2}|^2,
\end{eqnarray}
and
\begin{eqnarray}
\label{eq:chi12}
|\chi_{1,2}\rangle =  |\chi_{1,2}^{(0)}\rangle \pm  \frac{e^{\pm 2i \varphi}(\alpha-1)\kappa\Omega}{2(\Lambda + \Omega)\Lambda^2}|\psi_{1,2}|^2\,  |\chi_{2,1}^{(0)}\rangle.
\end{eqnarray} 
Here $|\psi_{1,2}|^2$ correspond to $|\psi_+|^2$ in the first and the second dressed states, respectively. 

Assuming that an eigenstate of $\hat{H}_0$ can be expressed as 
\begin{equation}
\xi = \psi_1 |\chi_1\rangle + \psi_2 |\chi_2\rangle,
\end{equation}
where $\psi_2$ remains small for all times,  the effective Hamiltonian describing the adiabatical dynamics of $\psi_1$ can be found as \cite{Malomed17S}
\begin{equation}
\label{eq:H1}
\hat{H}_1 = (\hat{p}-A)^2 + W + \Lambda + \frac{2\Omega^2 +\kappa^2(\alpha+1)}{\Lambda(\Lambda + \Omega)}|\psi_{1}|^2,
\end{equation}
where $A$ and $W$ are  geometric potentials \cite{Dalibard11S}
\begin{eqnarray}
A = i\langle \chi_1 | \partial_\varphi \chi_1\rangle, \quad W = |\langle \chi_2 | \partial_\varphi \chi_1\rangle|^2
\end{eqnarray}

Using the explicit form of the spinors, i.e., (\ref{eq:chi12}), we arrive at  the following density-dependent expressions:
\begin{eqnarray}
A &=& \left(\frac{\Omega}{\Lambda} -1\right)  + (1-\alpha)\frac{\Omega}{\Lambda^2} \left(1-\frac{\Omega}{\Lambda} \right)|\psi_1|^2,\\
W &=& \frac{\kappa^2}{\Lambda^2}   - \frac{(1-\alpha)\kappa^2\Omega}{\Lambda^4} |\psi_1|^2,
\end{eqnarray}
Then after dropping the irrelevant constant energy offset, we rewrite $\hat{H}_1$ in (\ref{eq:H1}) as 
\begin{equation}
\hat{H}_1 = (\hat{p}-A)^2 +g|\psi_1|^2,
\end{equation}
where the net nonlinearity coefficient reads
$$g = \frac{2\Omega^2 +\kappa^2(\alpha+1)}{\Lambda(\Lambda + \Omega)}  - \frac{(1-\alpha)\kappa^2\Omega}{\Lambda^4} .$$

\end{document}